\documentclass[12pt, prd, showpacs,showkeys,nofootinbib]{revtex4}
\usepackage{amssymb}
\usepackage{amsmath}

\input{tcilatex}

\begin{document}

\title{Acceleration of particles as a universal property of ergosphere}
\author{O. B. Zaslavskii}
\affiliation{Department of Physics and Technology, Kharkov V.N. Karazin National
University, 4 Svoboda Square, Kharkov 61022, Ukraine}
\email{zaslav@ukr.net }

\begin{abstract}
We show that recent observation made in Grib and Pavlov, \textit{Europhys.
Lett.} \textbf{101} (2013) 20004 for the Kerr black hole is valid in the
general case of rotating axially symmetric metric. Namely, collision of two
particles in the ergosphere leads to indefinite growth of the energy in the
centre of mass frame, provided the angular momentum of one of two particles
is negative and increases without limit for a fixed energy at infinity.
General approach enabled us to elucidate, why the role of the ergosphere in
this process is crucial.\bigskip
\end{abstract}

\maketitle

\section{Introduction}

In recent years, the interest to high-energetic processes near black holes
revived due to observation \cite{ban} made by Ba\~{n}ados, Silk, and West
(hereafter, BSW effect), so several dozens of papers appeared since 2009.
These authors found that under certain conditions, the energy $E_{c.m.}$ in
the centre of mass frame of two particles colliding near a black hole can
grow unbound. Meanwhile, potential attempts to observe the consequences of
the BSW effect are faced with difficulties since the strong red shift
restricts the energies of possible outcome which can be detected at infinity 
\cite{p} - \cite{z}. Quite recently, a new mechanism of getting ultra-high
energies in black hole physics was suggested in \cite{ergo}. Grib and Pavlov
demonstrated there that inside the ergosphere of the Kerr black hole, $%
E_{c.m.}$ also grows indefinitely if the angular momentum of one of
particles takes large negative values. Then, the aforementioned problems
with detecting at infinity do not arise since collision occurs, in general,
outside the horizon. For such an effect, there is no requirement to
fine-tune the parameters of one particle in contrast to the BSW one \cite%
{ban} that also facilitates the realization of the phenomenon under
discussion. It is worth stressing that interest to the potential role of the
ergosphere in the high energy processes revived recently in connection with
the problem of jet collimation \cite{jet1}, \cite{jet2}.

The main goal of the present work is to show that the effect of indefinite
growth of $E_{c.m.}$ occurs inside the ergosphere of a generic "dirty"
(surrounded by matter) stationary rotating axially symmetric black hole. We
also elucidate the role of the ergoregion in the effect under discussion.
This is done in a general form, without resorting to the properties of a
particular metric. In the original paper \cite{ban}, the BSW effect was
considered for extremal black holes and it took some efforts to show that a
similar effect can occur also near nonextremal ones \cite{gp}. For the
process in question \cite{ergo}, it is clear from the very beginning that a
black hole can be extremal or nonextremal. Moreover, as is clear from
consideration below, the horizon can be absent at all but the existence of
the ergosphere is mandatory (examples of spacetimes with the ergosphere but
without the horizon are discussed in \cite{gk}).

We use the systems of units in which the fundamental constants $G=c=1$.

\section{Form of metric and geodesics equations}

Let us consider the metric%
\begin{equation}
ds^{2}=-N^{2}dt^{2}+g_{\phi \phi }(d\phi -\omega dt)^{2}+\frac{\rho ^{2}}{%
\Delta }dr^{2}+g_{\theta \theta }d\theta ^{2}.  \label{m}
\end{equation}%
Here, the metric coefficients do not depend on $t$ and $\phi $. On the
horizon $N=0$. \ In (\ref{m}), the factor $\Delta (r)\sim N^{2}$ is singled
out for convenience. The coefficient $\rho $ can depend on $\theta $. The
Kerr metric belongs just to this class. We assume that, similarly to the
Kerr metric,%
\begin{equation}
\omega >0  \label{pos}
\end{equation}%
everywhere.

If \ a neutral particle of the mass $m$ moves freely in the spacetime (\ref%
{m}), its energy $E=-mu_{0}$ and the angular momentum $L=mu_{\phi }$ are
conserved due to the existence of Killing vectors responsible for time
translation and rotation. Here, $u^{\mu }=\frac{df^{\mu }}{d\tau }$ is the
four-velocity, $x^{\mu }=(t,\phi ,r,\theta )$. Then, the equators of motion
along geodesics read

\begin{equation}
m\dot{t}=mu^{0}=\frac{X}{N^{2}}\text{, }X=E-\omega L\text{.}  \label{t}
\end{equation}%
We assume the forward in time condition $\dot{t}>0$, so that 
\begin{equation}
E-\omega L\geq 0  \label{ft}
\end{equation}%
should be satisfied. Then, it follows from the geodesic equations and the
normalization condition $u_{\mu }u^{\mu }=-1$ that%
\begin{equation}
m\dot{\phi}=\frac{L}{g}+\frac{\omega X}{N^{2}}\text{, }g=g_{\phi \phi }
\label{phi}
\end{equation}%
that can be also rewritten as%
\begin{equation}
m\dot{\phi}=\frac{\omega E}{N^{2}}-L\frac{g_{00}}{gN^{2}}\text{,}  \label{pl}
\end{equation}%
\begin{equation}
\frac{\rho }{\sqrt{\Delta }}m\dot{r}=\varepsilon \frac{Z}{N},  \label{r}
\end{equation}%
where dot denotes the derivative with respect to $\tau $. Here, 
\begin{equation}
Z^{2}=X^{2}-N^{2}(m^{2}+\frac{L^{2}}{g}+m^{2}g_{\theta \theta }\dot{\theta}%
^{2})\text{,}  \label{Z}
\end{equation}%
$\varepsilon =sign\dot{r}$.

Using the relation%
\begin{equation}
g_{00}=-N^{2}+\omega ^{2}g\text{,}
\end{equation}

we can rewrite eq. (\ref{Z}) as%
\begin{equation}
Z^{2}=\frac{L^{2}g_{00}}{g}-2E\omega L+E^{2}-N^{2}m^{2}(1+B)\text{, }%
B=g_{\theta \theta }\dot{\theta}^{2}\geq 0\text{,}  \label{z1}
\end{equation}%
or%
\begin{equation}
Z^{2}=\frac{g_{00}}{g}(L_{-}-L)(L_{+}-L),  \label{zl}
\end{equation}%
where $L_{+}$ and $L_{-}$ are the roots of the equation $Z=0$. According to (%
\ref{r}), this corresponds to turning points $\dot{r}=0$ . For the Kerr
metric, the variables in the equations for geodesics are separated, and the
values $L_{+}$ and $L_{-}$ can be expressed in terms of the Carter constant
- see eqs. (10) and (18) of \cite{ergo}. For generic dirty black holes, the
roots $L_{+}$ and $L_{-}$ depend on $\dot{\theta}$ and cannot be found in
the closed form in general. This can be done for equatorial motion, $\dot{%
\theta}=0=B$. Then, we find from (\ref{zl}) 
\begin{equation}
L_{\pm }=\frac{g\omega }{g_{00}}E\pm \frac{N\sqrt{g}}{g_{00}}\sqrt{%
E^{2}+m^{2}g_{00}}\text{.}  \label{+}
\end{equation}

\section{Ergosphere and integrals of motion}

By definition, the surface of the ergosphere is defined by equation $%
g_{00}=0 $. We will discuss some general properties of three regions
separately. For what follows, we also need the asymptotic expressions for $Z$
for large $\left\vert L\right\vert $. It is clear from (\ref{pos}) and (\ref%
{ft}) that the limit $\left\vert L\right\vert \rightarrow \infty $ with
fixed $E$ can be realized for $L=-\left\vert L\right\vert <0$ only.

\subsection{Outside ergosphere, $g_{00}<0$}

Then, it follows from (\ref{z1}), the condition $Z^{2}\geq 0$ and (\ref{ft})
that%
\begin{equation}
E\geq \omega L+\sqrt{\omega ^{2}L^{2}+N^{2}m^{2}(1+B)+\frac{L^{2}\left\vert
g_{00}\right\vert }{g}}\geq 0\text{,}  \label{epos}
\end{equation}%
In terms of angular momenta, $L_{+}\leq L\leq L_{-}.$ It is seen from (\ref%
{z1}) that for $L\rightarrow -\infty $ and $E$ fixed, the first term is
negative and dominates $Z^{2}$, so this limit cannot be realized in this
region.

\subsection{Boundary of ergosphere, $g_{00}=0$}

Then, it follows from (\ref{pl}) that $m\dot{\phi}=\frac{\omega E}{N^{2}}$
does not depend on $L$ for a fixed energy. In (\ref{zl}), the terms of the
order $L^{2}$ in $Z^{2}$ cancel. For $\left\vert L_{2}\right\vert
\rightarrow \infty $ it turns out that 
\begin{equation}
Z\approx \sqrt{2E\omega \left\vert L\right\vert },  \label{zb}
\end{equation}%
where $E\geq 0$ according to (\ref{epos}).

\subsection{Inside ergosphere, $g_{00}>0$}

If $L=-\left\vert L\right\vert <0$, the second term in (\ref{pl}) is
positive. Thus for a fixed energy the angular velocity of rotation is
increasing function of $\left\vert L\right\vert $. The more negative becomes
the angular momentum, the larger the angular velocity of rotation in the
positive direction! This generalizes observation made in \cite{ergo} for the
Kerr metric. For large $\left\vert L\right\vert $,%
\begin{equation}
Z=\left\vert L\right\vert \sqrt{\frac{g_{00}}{g}}+O(1)\text{.}  \label{z00}
\end{equation}

Thus we see that trajectories with fixed $E$ and $L=-\left\vert L\right\vert
\rightarrow -\infty $ are possible inside the ergosphere or on its boundary
but are forbidden outside it. This generalizes the corresponding property of
the Kerr metric \cite{ergo}.

\section{Energy in centre of mass frame}

For two colliding particles, in the point of collision one can define the
energy in the centre of mass frame as%
\begin{equation}
E_{c.m.}^{2}=-\left( m_{1}u_{1}^{\mu }+m_{2}u_{2}^{\mu }\right) \left(
m_{1}u_{1\mu }+m_{2}u_{2\mu }\right)
\end{equation}%
that is the counterpart of the standard textbook formula for one particle $%
m^{2}=-p^{\mu }p_{\mu }$ where $p^{\mu }\,$\ is the four-momentum.

Then,%
\begin{equation}
E_{c.m.}^{2}=m_{1}^{2}+m_{2}^{2}+2m_{1}m_{2}\gamma  \label{e}
\end{equation}%
where the Lorentz factor of relative motion%
\begin{equation}
\gamma =-u_{1}^{\mu }u_{2\mu }\text{,}  \label{gu}
\end{equation}%
$\left( u^{\mu }\right) _{i}$ is the four-velocity of the i-th particle
(i=1,2). Using (\ref{t}) - (\ref{r}) we obtain from (\ref{gu}) that%
\begin{equation}
\gamma =\frac{1}{m_{1}m_{2}}(c-d)--g_{\theta \theta }\dot{\theta}_{1}\dot{%
\theta}_{2}\text{, }c=\frac{X_{1}X_{2}-\varepsilon _{1}\varepsilon
_{2}Z_{1}Z_{2}\text{, }}{N^{2}}\text{, }d=\frac{L_{1}L_{2}}{g}.
\label{gamma}
\end{equation}

\section{Collisions with large negative angular momentum}

We are interested in the situation when $E_{c.m.}^{2}$ can be made as large
as possible. One possibility is to make denominator in $c$ (\ref{gamma})
small due to $N\rightarrow 0$. This corresponds to the BSW effect.
Meanwhile, there is one more option to which Grib and Pavlov paid attention 
\cite{ergo} - the case of large numerator. Let $L_{1}$ be finite but the
absolute value of $L_{2}$ be very large (by analogy with the BSW effect, we
call particle 1 usual and particle 2 critical). As is explained in Sec. III,
this is only possible if (i) $L_{2}=-\left\vert L_{2}\right\vert <0$, (ii)
collisions occur inside the ergosphere or near its boundary.

\subsection{Collisions inside ergosphere}

In the limit $\left\vert L_{2}\right\vert \rightarrow \infty $, using (\ref%
{zl}) and retaining in (\ref{gamma}) the leading terms in $L_{2}$, we obtain%
\begin{equation}
E_{c.m.}^{2}=\frac{2\left\vert L_{2}\right\vert }{N^{2}}\alpha +O(1)\text{,}
\label{e2}
\end{equation}%
\begin{equation}
\alpha =\omega X_{1}-\varepsilon _{1}\varepsilon _{2}Z_{1}\sqrt{\frac{g_{00}%
}{g}}+\frac{L_{1}N^{2}}{g}>0  \label{al}
\end{equation}%
Taking into account the useful relation%
\begin{equation}
L_{+}+L_{-}=\frac{2E}{g_{00}}g\omega
\end{equation}%
that follows from (\ref{z1}), one can check that $2\alpha =\frac{g_{00}}{g}%
[\varepsilon _{1}\sqrt{\left( L_{1}\right) _{+}-L_{1}}-\varepsilon _{2}\sqrt{%
\left( L_{1}\right) _{-}-L_{1}}]^{2}$, whence%
\begin{equation}
E_{c.m.}^{2}\approx \frac{2\left\vert L_{2}\right\vert g_{00}}{N^{2}g}%
[\varepsilon _{1}\sqrt{\left( L_{1}\right) _{+}-L_{1}}-\varepsilon _{2}\sqrt{%
\left( L_{1}\right) _{-}-L_{1}}]^{2}\text{.}  \label{en}
\end{equation}%
It is seen from (\ref{en}) that it is the property $g_{00}>0$ inside the
ergosphere that ensures the positivity of $E_{c.m.}^{2}$.

\subsection{Collisions near the boundary of ergosphere}

On the boundary of the ergosphere $g_{00}=0$, the approximate expressions (%
\ref{e2}), (\ref{al}) are not valid. Using (\ref{zb}), one obtains

\begin{equation}
E_{c.m.}^{2}=\frac{2}{N^{2}}(\omega E_{1}\left\vert L_{2}\right\vert
-Z_{1}\varepsilon _{1}\varepsilon _{2}\sqrt{2E_{2}\omega \left\vert
L_{2}\right\vert })+O(1).  \label{nb}
\end{equation}

Thus inside the ergosphere and on the boundary the leading term has the
order $\left\vert L_{2}\right\vert $ but the subleading terms are different:
\ they are finite inside and proportional to $\sqrt{\left\vert
L_{2}\right\vert }$ on the boundary.

\section{Case of circular orbits}

There is also a special case when $\theta =\frac{\pi }{2}$ and $L_{2}$ is
adjusted to ensure the circular character of the orbit on which $\dot{r}%
_{2}=0=Z_{2}$. Then, we have from (\ref{e}) and (\ref{gamma}) that%
\begin{equation}
E_{c.m.}^{2}=m_{1}^{2}+m_{2}^{2}+2\left( \frac{X_{1}X_{2}}{N^{2}}-\frac{%
L_{1}L_{2}}{g}\right) \text{.}
\end{equation}

Let us assume that we are outside the ergosphere, so in our notations (which
are here opposite to those in \cite{ergo}) $L_{2}=\left( L_{2}\right)
_{+}<\left( L_{2}\right) _{-}$ $.$ If we choose the radius of the orbit
closer and closer to the boundary of the ergosphere, $g_{00}\rightarrow -0$.
Then, it follows from (\ref{+}) that 
\begin{equation}
L_{2}\approx -\frac{2NE_{2}\sqrt{g}}{\left\vert g_{00}\right\vert }%
\rightarrow -\infty ,  \label{l+}
\end{equation}%
whence%
\begin{equation}
E_{c.m.}^{2}\approx -\frac{4E_{1}E_{2}\sqrt{g}\omega }{g_{00}N}  \label{be}
\end{equation}%
diverges as the boundary is approached. It is worth noting that the leading
terms in (\ref{nb}) and (\ref{be}) coincide but the subleading ones are
different since $Z_{2}$ is large in the first case and vanishes in the
second one.

Another case of interest arises when particle 2 is orbiting with $%
E_{2}=-\left\vert E_{2}\right\vert $ from inside. Then, in a similar way, $%
L_{2}=L_{-}$ $\approx -\frac{2N\left\vert E_{2}\right\vert \sqrt{g}}{%
\left\vert g_{00}\right\vert }$. For particle 1, \thinspace $L_{1}$ is
finite and the energy obeys inequality (\ref{epos}) in which the term with $%
g_{00}$ is neglected. Then, eq. (\ref{be}) is valid in which, however, now $%
g_{00}\rightarrow +0$.

\section{Comparison with the Kerr metric}

In this case, in the Boyer-Lindquiste coordinates,%
\begin{equation}
\omega =\frac{2aMr}{(r^{2}+a^{2})^{2}-a^{2}\Delta \sin ^{2}\theta }\text{,}
\end{equation}%
\begin{equation}
g=\frac{\sin ^{2}\theta }{\rho ^{2}}[(r^{2}+a^{2})^{2}-a^{2}\Delta \sin
^{2}\theta ],
\end{equation}%
\begin{equation}
N^{2}=\frac{\rho ^{2}\Delta }{(r^{2}+a^{2})^{2}-a^{2}\Delta \sin ^{2}\theta }%
\text{,}
\end{equation}%
\begin{equation}
g_{00}=-(1-\frac{2Mr}{\rho ^{2}})\text{,}
\end{equation}%
\begin{equation}
\rho ^{2}=r^{2}+a^{2}\cos ^{2}\theta ,\text{ }\Delta =r^{2}-2Mr+a^{2}\text{.}
\end{equation}

Here, $M$ is the black hole mass, $a=\frac{J}{M}$ where $J$ is the black
hole angular momentum. Then, one can check that eq. (\ref{en}) turns into
eq. (25) of \cite{ergo}. It is worth noting that the asymptotic form (\ref%
{en}) is valid independently of special symmetry properties of the metric
which were used in \cite{ergo} for the Kerr case.

\section{Kinematic explanation}

The indefinitely growing energy in the centre of mass frame implies that the
relative velocity tends to that of light. This occurs when one of particles
has the velocity close to that of light whereas the other one moves with
some finite velocity. For the BSW effect, such an explanation was given in 
\cite{k}. It is based on the relationship (see eq. (15) of \cite{k})

\begin{equation}
E-\omega L=\frac{mN}{\sqrt{1-V^{2}}}\text{,}  \label{kin}
\end{equation}%
where $V$ is the local velocity of a particle in the zero angular momentum
frame \cite{bar}. In the horizon limit $N\rightarrow 0$, the velocity $%
V_{us}\rightarrow 1$ if $E-\omega L>0$ does not vanish on the horizon (a
usual particle) and $V_{cr}<1$ if $E-\omega L=0$ on the horizon (the
critical one).

Now, the situation is different since inside the ergosphere or on its
boundary $N\neq 0$ in general (the pole points are exception but in their
vicinity the ergosphere approaches the horizon, so standard BSW effect takes
place). Instead, now for the critical particle the second term in the left
hand side of (\ref{kin}) tends to infinity, so that the right hand side must
also diverge, whence $V_{cr}\rightarrow 1$. For a usual particle, $V$ takes
some finite value $V_{us}<1$. Thus instead of $V_{cr}<1$, $V_{us}\rightarrow
1$ for the BSW effect, now, vice versa, $V_{cr}\rightarrow 1$ and $V_{us}<1$
for collisions in or on the ergosphere.

The case under discussion possesses two interesting features: (i) $%
\left\vert L_{2}\right\vert \rightarrow \infty $ but $E_{2}$ is finite, (ii)
this effect happens inside the ergosphere or on its boundary and is
impossible outside it.

\section{Conclusion}

Thus we suggested model-independent approach and generalized an interesting
observation by Grib and Pavlov \cite{ergo} to arbitrary "dirty" (surrounded
by matter) rotating axially symmetric black holes. Some relevant properties
of motion of geodesic particles are elucidated depending on their angular
momentum and its location (outside, inside or on the boundary of the
ergosphere). We also gave general explanation of the role of the ergosphere
in the process of collisions with indefinite growth of $E_{c.m.}$ due to
large negative angular momentum of the critical particle. As is clear from
this approach, such an effect occurs even without black hole horizons,
provided the ergosphere exists.

Thus the possibility of acceleration of particles to ultra-high energies
turns out to be a universal property of the ergosphere.

\begin{acknowledgments}
I\ thank Yurii Pavlov for stimulating correspondence. 
\end{acknowledgments}

\end{document}